# *PHONONICS* IN LOW-DIMENSIONS: ENGINEERING PHONONS IN NANOSTRUCTURES AND GRAPHENE

Alexander A. Balandin[a,*] and Denis L. Nika[a,b]

[a]*Department of Electrical Engineering and Materials Science and Engineering Program, Bourns College of Engineering, University of California, Riverside, CA 92521 U.S.A.*

[b]*E. Pokatilov Laboratory of Physics and Engineering of Nanomaterials, Department of Theoretical Physics, Moldova State University, Chisinau, MD-2009, Republic of Moldova*

[c]*e-mail: balandin@ee.ucr.edu*

**Abstract**

Phonons – quanta of crystal lattice vibrations – reveal themselves in all electrical, thermal and optical phenomena in materials. Nanostructures open exciting opportunities for tuning the phonon energy spectrum and related properties of materials for specific applications. The possibilities for controlled modification of the phonon transport and phonon interactions – referred to as *phonon engineering* or *phononics* – increase even further with the advent of graphene and two-dimensional van der Waals materials. We describe methods for tuning the phonon spectrum and controlling the thermal properties of the low-dimensional materials via ribbon edges, grain boundaries, isotope composition, defect concentration, and atomic-plane orientation.

# PHONONS AND PHONON ENGINEERING

Phonons, i.e. quanta of the crystal lattice vibrations, affect all physical processes in solids [1]. They limit the electron mobility near room temperature (RT), and affect optical properties of crystalline materials. Acoustic phonons are the main heat carriers in insulators and semiconductors. The long-wavelength phonons give rise to sound waves, which explains the name *phonon*. Similar to electrons, phonons are characterized by their dispersion $\omega(q)$, where $\omega$ is an angular frequency and $q$ is a wave vector of a phonon. In bulk semiconductors with $g$ atoms per unit cell, there are $3g$ phonon dispersion branches for each $q$. Three types of vibrations describe the motion of the unit cell, and form three acoustic phonon branches. The other $3(g-1)$ modes describe the relative motion of atoms inside a unit cell, and form the optical phonon branches. The acoustic polarization branches are commonly referred to as longitudinal acoustic (LA) and transverse acoustic (TA). In case of graphene the out-of-plane transverse vibrations are demoted as z-axis acoustic (ZA) phonons. In the long-wavelength limit, acoustic phonons in bulk crystals have nearly linear dispersion, which can be written as $\omega = V_S q$, where $V_S$ is the sound velocity, while the optical phonons are nearly dispersion-less and have a small group velocity $V_G = d\omega/dq$.

## Spatial Confinement and Localization of Phonons

Spatial confinement of acoustic phonons in nanostructures affects their dispersion [2-5]. It modifies acoustic phonon properties such as phonon group velocity, polarization, density of states, and changes the way acoustic phonons interact with other phonons, defects and electrons [4-5]. Such changes create opportunities for engineering phonon spectrum in nanostructures for improving electrical or thermal properties. The average phonon mean free path (MFP) $\Lambda$ in semiconductors is ~50-300 nm near RT. The wavelength of the thermal phonon $\lambda_0 = 1.48 \hbar V_S / (k_B T)$ is ~1-2 nm ($\hbar$ is Planck's constant, $k_B$ is the Boltzmann constant, $T$ is absolute temperature) [6]. Thus, in order to engineer the acoustic phonon spectrum at RT, one needs to have materials structured at the nanometer length-scale. Fig. 1 presents phonon dispersion in three-dimensional (3-D) bulk, a semiconductor thin film and a nanowire. Embedding nanostructures in materials with the large acoustic impedance mismatch, gives one greater flexibility for tuning the phonon spectrum [4-5, 7-10]. The acoustic impedance is defined as $\zeta = \rho V_s$, where $\rho$ is the mass density [7]. In the acoustically mismatched nanowires the phonon dispersion inside the nanowires depends not only on its diameter and the boundary conditions at the nanowires surface but also on the material of the barrier layers [4-5, 8-9].

Engineering of the optical phonons in nanostructures via the boundary conditions requires different approaches than engineering of the acoustic phonons. In the long-wave limit the optical phonons correspond to the motion of atoms within the same unit cell, which cannot be changed by imposing new outside boundaries. However, the electron – phonon scattering rates can be modified by tuning the confined electronic states energy difference with respect to the optical phonon energy [11]. This effect – referred to as "phonon bottleneck" – can be used for optimization of solid-state lasers or other devices. Heterostructures, which consist of the layers with the distinctively different optical phonon energies, allow one to localize optical phonons within their respective layers [1, 12-13], which can also be used for practical purposes.

Although phonon engineering became a mainstream research direction only recently, the interest to modification of the acoustic phonon spectra in layered materials has a long history. In 1950s, the changes in acoustic vibrations leading to appearance of folded phonons were analyzed in the "artificial thinly-laminated media" – structures, which now would be called superlattices [14]. The folded phonons were later observed experimentally in GaAs/AlGaAs quantum well superlattices [15]. In 1990s, many calculations were performed for the confined acoustic phonon – electron scattering rates in freestanding thin films and nanowires [16-19].

The interest to the subject substantially increased when it was pointed out that the confinement-induced changes in the acoustic phonon dispersion can lead to strong effects on thermal conductivity [2-3]. Decreased averaged phonon group velocity in thin films and nanowires can lead to the increased acoustic phonon relaxation on point defects (vacancies, impurities, isotopes, etc.), dislocations and phonon-phonon Umklapp processes [20]. Thermal conductivity reduction, being a bad news for thermal management of downscaled electronic devices, is good news for the thermoelectric devices, which require materials with high electrical conductivity, Seebeck coefficient and low thermal conductivity [2, 21].

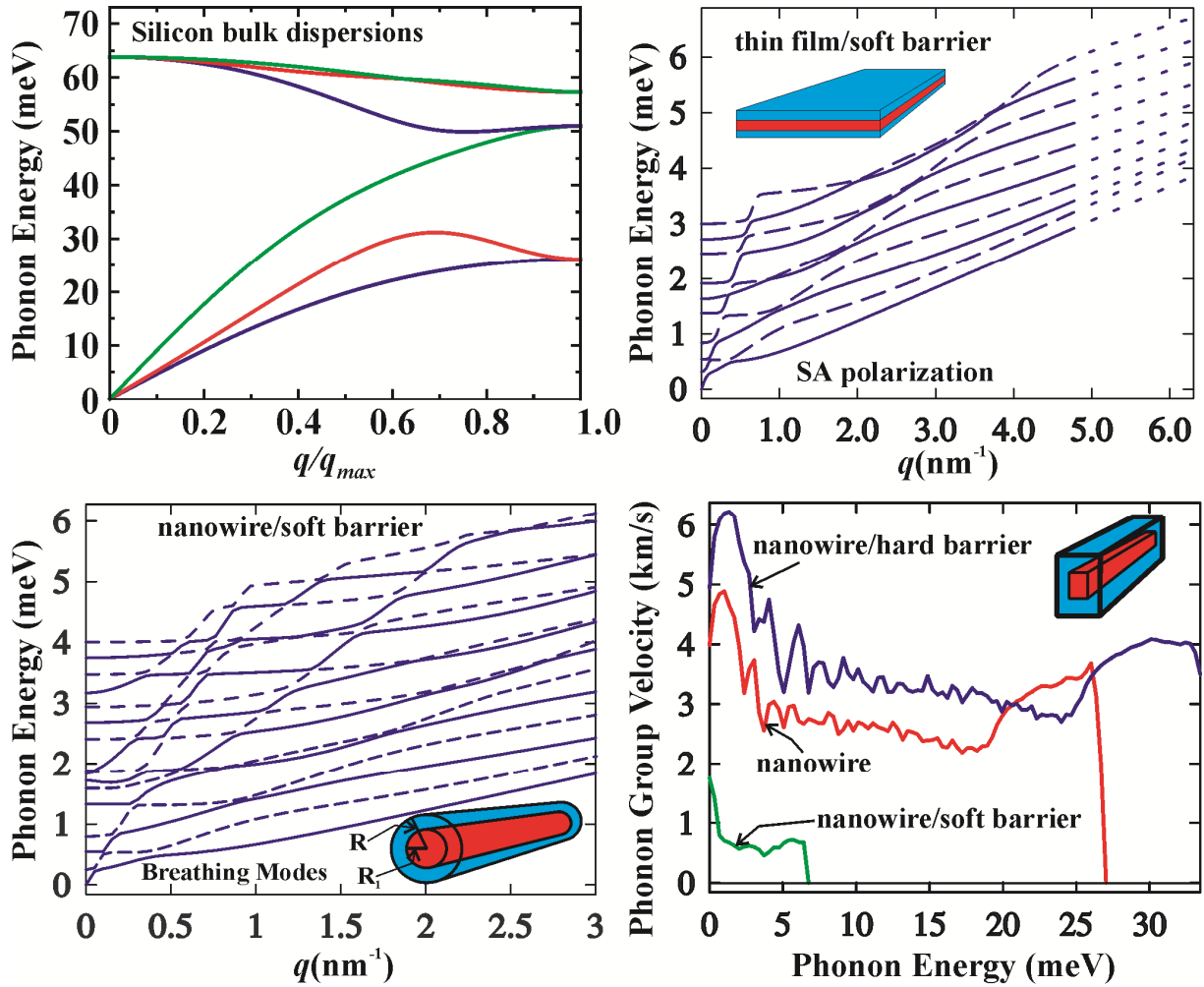

**Figure 1:** Phonon dispersion engineering in semiconductor nanostructures. (a) Phonon energy dispersion in a bulk silicon crystal along [1,1,0] crystallographic direction calculated using the five-parameter Born-von Karman model. (b-c) Phonon energy dispersions in wurtzite GaN film (b) and GaN nanowires (c) covered with the soft barrier material. (d) Phonon group velocities averaged over all phonon branches in generic GaN nanowire and GaN nanowires covered with the acoustically hard wurtzite AlN and soft plastic barriers. Note strong changes in the acoustic phonon dispersion and group velocities, which affect phonon interactions with other phonons, defects and electrons. Figure is adopted from Refs. [7-9] with permission from the American Institute of Physics and American Physical Society.

**Phonon Thermal Conductivity and Phonon-Limited Electron Mobility**

When the structure size $L$ is comparable to the phonon MFP but still much larger that the dominant thermal phonon wavelength, the phonon dispersion remains bulk-like. In this case, the acoustic phonon transport is only affected by phonon scattering from the boundaries. The phonon boundary scattering rate can be evaluated as [22] $1/\tau_B=(V_S/D)[(1-p)/(1+p)]$, where $D$ is the nanostructure size and $0 \leq p \leq 1$ is the specularity parameter defined as a probability of specular and diffuse scattering at the boundary. In nanostructures, where the phonon - boundary scattering is dominant, thermal conductivity scales with the size $D$ as $K_p \sim C_p V_S \Lambda \sim C_p V_S^2 \tau_B \sim C_p V_S D$, where $C_p$ is the thermal heat capacity at the constant pressure.

The situation for the phonon thermal conductivity and phonon-limited electron transport becomes much more interesting when $L$ becomes comparable to $\lambda_0$. In this case, the spatial confinement of acoustic phonon and mode quantization open an opportunity for increasing or decreasing the thermal conductivity and electron mobility via engineering the phonon spectrum. It has been known since 1980 that the electron mobility limited by the elastic scattering, e.g. ionized impurity scattering, can be strongly increased in nanowires via the restriction of the scattering space available for electrons in a quasi-1-D system [23]. However, the RT electron mobility in semiconductor crystals is limited by phonons rather than impurities. Recently, it was shown theoretically that the electron mobility limited by the phonons in Si nanowires [24] or thin films [25] can be enhanced via suppression of electron-phonon interactions in nanostructures with synthetic diamond barriers (see Fig. 2). The acoustically hard diamond barriers result in the modification of the phonon dispersion inside Si channel layer beneficial for the electron transport, e.g. increasing charge carrier mobility [24-25]. Similarly, one can increase or decrease the heat conduction properties of the nanowire or thin film by using the proper acoustically hard or soft boundaries [26-27]. One should note here that the prediction for thermal and electronic conduction in semiconductor nanostructures in the phonon confinement regime initially made within the elastic continuum approximation [2-5, 7-10, 20-21, 24-34] have been confirmed by the independent molecular-dynamics simulations [35] and direct experimental measurements for Ge-Si core-shell nanowires [36]. The described phonon engineering approach can be used in the electronic industry for design of nanoscale transistors and phononic band gap materials [28-34]. As the transistor feature size approaches $\lambda_0$ the possibilities for engineering phonon dispersion to improve the carrier and heat transport increase, correspondingly.

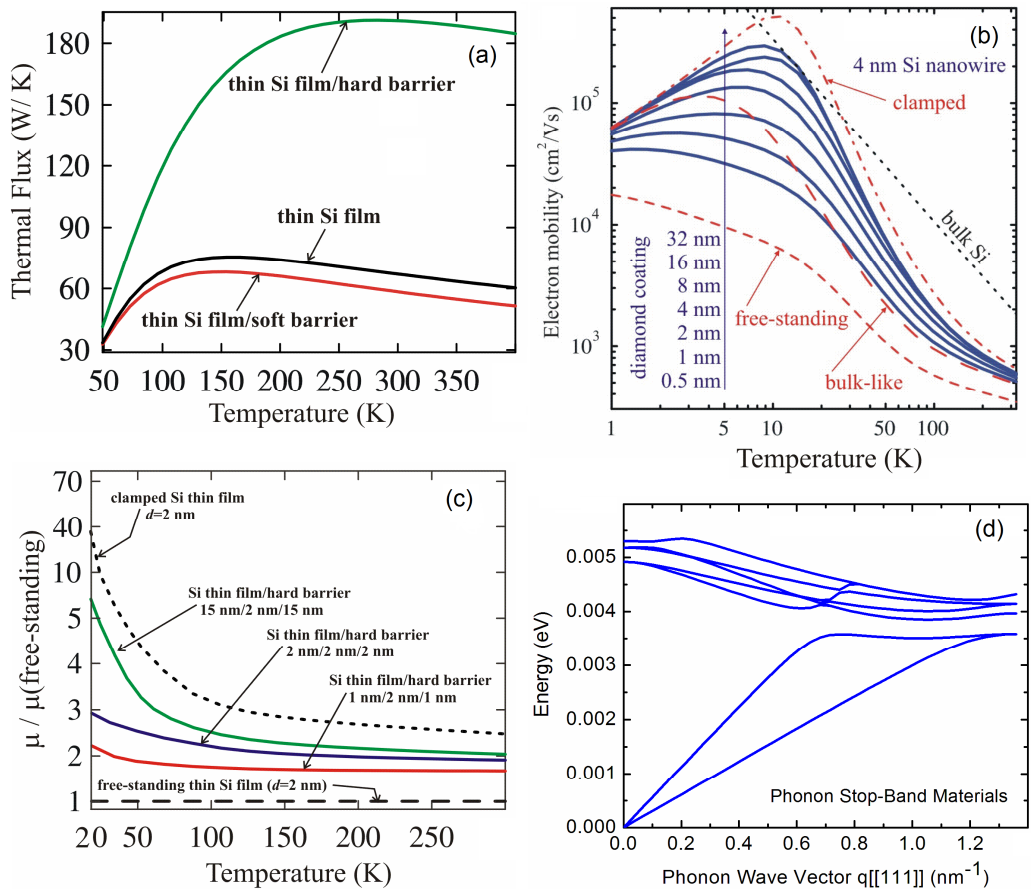

**Figure 2:** Modification of thermal conductivity and electron mobility via phonon engineering in nanostructures. (a) Temperature dependence of phonon heat flux per temperature gradient and width unit in 4.88-nm thick Si thin film and three-layered heterostructures diamond/Si/diamond (3 nm/ 4.88 nm/ 3 nm) and plastic/Si/plastic (3 nm/ 4.88 nm/ 3 nm). (b-c) Temperature dependence of electron mobility in Si nanowires and thin without barriers and in nanowires and thin films covered with the hard diamond barriers. (d) Phonon stop band in ordered array of semiconductor quantum dots. The phonon stop band results in reduction of the thermal conductivity beneficial for thermoelectric applications. Figures (a-c) are adopted from Refs. [24-25, 27] with permission from the American Chemical Society, American Institute of Physics and American Scientific Publishers. Figure (d) is based on the data reported in Ref. [28].

**PHONONS AND THERMAL PROPERTIES OF GRAPHENE**

Acoustic phonons are the main heat carriers in carbon materials. Although graphite reveals many metal characteristics, its heat transport is dominated by phonons owing the exceptionally strong sp$^2$ covalent bonding of its lattice. The thermal conductivity of various allotropes of carbon span an extraordinary large range — of over five orders of magnitude — from ~0.01 W/mK in amorphous carbon to above 2000 W/mK in diamond or graphite at RT [37]. In 2007, the first measurements of the thermal conductivity of graphene carried out at UC Riverside revealed unusually high values of thermal conductivity. The values measured for the high-quality large suspended graphene samples (length above 10 μm) were exceeding those for basal planes of graphite [38-39].

The experimental observation was explained theoretically by the specifics of the 2-D phonon transport [40-42]. The low-energy acoustic phonons in graphene, which make substantial contribution to heat conduction, have extraordinary large MFP [39]. The anharmonic scattering in 2-D graphene is very weak for such phonons. The large values of thermal conductivity and 2-D phonon density of states make graphene an ideal material for phonon engineering. The thermal transport in 2-D graphene can be tuned more readily than in the corresponding bulk 3-D graphite [43-44]. The phonon heat conduction in thin films of the *van der Waals materials* – materials with the layered crystalline structure such as graphite or Bi$_2$Te$_3$ – can be less prone to the phonon scattering from the top and bottom interfaces owing to a possibility of obtaining very smooth interfaces (e.g. $p\approx1$).

**Experimental Studies of Thermal Transport in Graphene**

The experimental studies of thermal properties of graphene were made possible with development of the optothermal Raman technique (Fig. 3). In this technique, a Raman spectrometer acts as a thermometer measuring the local temperature rise in graphene in response to the laser heating [38-39, 45-47]. The measured thermal conductivity values for suspended graphene were in the range ~2000 – 5000 W/mK and depended strongly on the sample size and quality [38-39, 45]. Independent measurements conducted by other research groups using modified optothermal Raman techniques reveled the values in the range 1500 – 5000 W/mK at RT [48-49]. The thermal conductivity of the supported graphene was lower than that in suspended: $K \approx 600$ W/mK for graphene-on-SiO$_2$/Si near RT [50]. Encasing graphene within two layers of SiO$_2$ can lead to further reduction of the thermal conductivity due to the phonon interface and disorder scattering [51]. The thermal conductivity data for graphene is summarized in Table I. The $K$ values are given for room temperature unless indicated otherwise.

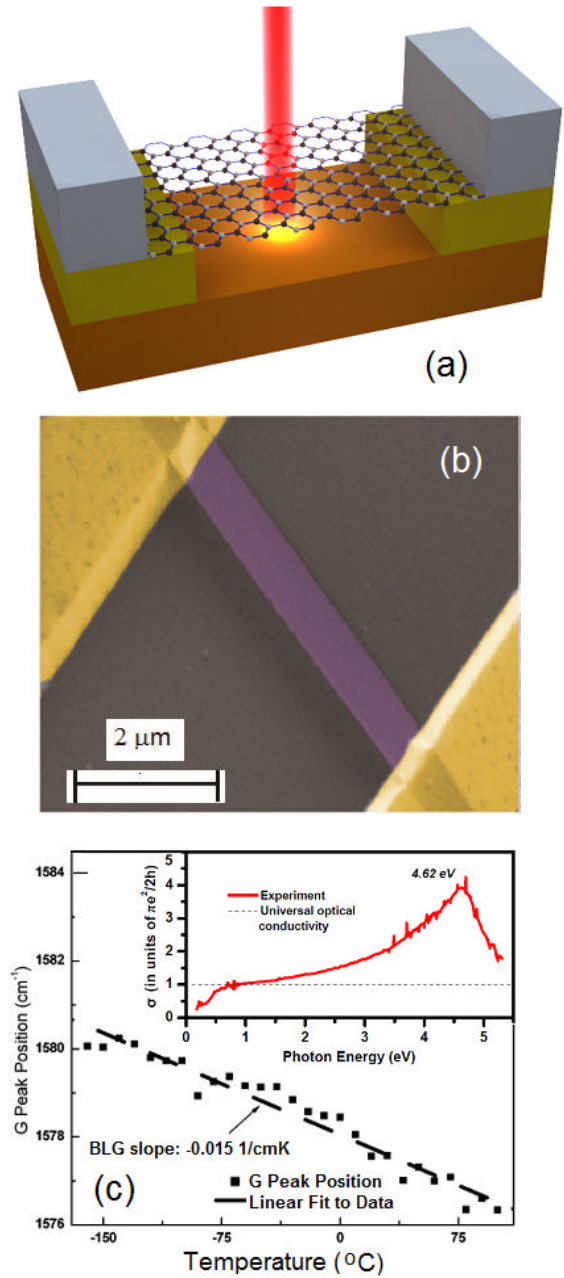

**Figure 3:** Optothermal measurements of the thermal conductivity of graphene. (a) Schematic of the experimental setup with the excitation laser light focused on graphene suspended across a trench in a Si wafer. Laser power absorbed in graphene induces a local hot spot and generates a heat wave propagating toward the heat sinks. (b) Pseudo-color scanning electron microscopy image of graphene flake suspended over trench. (c) Experimental data for Raman G-peak position as a function of laser power, which determines the local temperature rise in response to the dissipated power. Determining of the thermal conductivity requires measurements of light absorption in graphene at given excitation laser energy under conditions of the experiment. The inset shows experimental optical conductivity of monolayer graphene as the function of the photon energy in the spectral range of 0.2 – 5.5 eV after Ref. [65]. Figure is adopted from Refs. [37, 45, 58] with permission from the Deutsche Physikalische Gesellschaft and Nature Publishing Group.

**Table I**: Thermal conductivity of graphene and graphene nanoribbons

| Sample | Method | K(W m$^{-1}$K$^{-1}$) | Comments | Refs |
|---|---|---|---|---|
| graphene | Raman optothermal | ~ 2000 – 5000 | suspended, exfoliated | 38, 39 |
| | | ~ 2500 | suspended, CVD | 48 |
| | | ~ 1500 – 5000 | suspended, CVD | 49 |
| | | ~ 600 | suspended, exfoliated; $T$ ~ 660 K | 52 |
| | electrical | ~600 | supported, exfoliated | 50 |
| graphene | theory: BTE + RTA | 1000 – 8000 | average Gruneisen parameters; strong size dependence | 41 |
| | | 2000 – 8000 | mode-dependent Gruneisen parameters; strong width, edge, lattice defect and Gruneisen parameters dependence; Valence Force Field model of lattice dynamics | 40 |
| | theory: BTE + 3$^{rd}$-order interatomic force constants (IFCs) | ~ 2430 | $K$(graphene)>$K$(carbon nanotubes) | 53 |
| | | 1500 - 3500 | strong size dependence; optimized Tersoff interatomic potential | 42 |
| | theory: MD | 500 – 1100 | $T$ ~ 435 $K$, calculation domain 4.4 nm × 4.3 nm × 1.6 nm, periodic boundary conditions; optimized Tersoff interatomic potential (IP) | 54 |
| | theory: MD + Green-Kubo | 2900 | strong dependence on a vacancy concentration; reactive empirical bond-order carbon potential; | 55 |
| | theory: MD | ~ 2000 | T ~ 400 K, 1.5 nm × 5.7 nm zigzag GNR; strong edge and lattice defect dependence; Brenner IP | 56 |
| | theory: MD + Green-Kubo | ~ 2700 | original Tersoff IP | 57 |
| | | ~ 2100 | reoptimized Tersoff IP | |
| few-layer graphene | Raman optothermal | 2800 - 1300 | suspended, exfoliated; $n$ = 2 - 4 | 58 |
| | theory: BTE + RTA | 4000 – 900 | $N$=1-8, strong edge dependence; Valence Force Field model of lattice dynamics | 58 |
| | theory: BTE + 3$^{rd}$-order IFCs, | 3500 – 1000 | $N = 1 – 5$, strong size dependence; optimized Tersoff IP for in-plane interaction and Lennard-Jones IP for interplanar bonding | 59 |
| | | 3300 - 2000 | $N = 1 – 4$; optimized Tersoff IP for | 60 |

| | | | in-plane interaction and Lennard-Jones IP for interplanar interaction | |
|---|---|---|---|---|
| | theory: MD | 880 - 580 | $N = 1 - 5$, strong dependence on the Van-der-Waals bond strength; Tersoff IP for in-plane interaction and Lennard-Jones IP for interplanar interaction | 61 |
| few-layer graphene nanorribons | electrical self-heating | 1100 | supported, exfoliated; $n<5$ | 62 |
| graphene nanoribbons | theory: MD | 1000 - 7000 | strong ribbon width and edge dependence; Tersoff IP | 63 |
| | theory: BTE + RTA | ~ 5500 | nanoribbon with width 5 μm; strong dependence on the edge roughness | 64 |
| | theory: MD + Green-Kubo | 400 - 2300 | zigzag nanoribbons, strong lattice defect dependence | 57 |

**Acoustic Phonon Transport in Graphene Nanoribbons**

The phonon transport and corresponding thermal properties undergo substantial modification in graphene nanoribbons. In nanoribbons with the feature sizes smaller than phonon MFP, the phonon transport is ballistic and limited by the ribbon size and edge characteristics. The thermal conductivity usually increases with the size of the ribbon. The phonon transport was studied computationally, using MD [54-56, 61, 63, 66-69], Green's function method [70-73] and Boltzmann transport equation (BTE) [40-44, 53, 59-60, 64]. In BTE approach, RT thermal conductivities $K \sim 1000 - 8000$ W/mK were predicted for graphene flakes depending on the size, edge qualities and Gruneisen parameters, which characterizes the anharmonicy of the lattice [40-44, 53, 59-60, 64]. MD simulations gave $K \sim 500 - 3000$ W/mK at RT [54, 61]. The phonon spectrum of graphene has been determined from the continuum model [74-75], generalized gradient approximation [76-77], local density function approximation [76, 78], and different models of lattice vibrations [40, 59-60, 77, 79-81]. The difference in the calculated and measured [82-83] phonon energies can reach 50 – 60 meV and lead to discrepancy in the predicted thermal conductivity values (see Table I). One should note here that the question of what type of phonons makes the dominant contribution to heat conduction is still actively debated [37]. The argument for LA and TA phonons is based on observation that these phonons have much smaller Gruneisen parameter, which determines the strength of scattering, and higher group velocity [43-44]. The alternative view of large ZA contribution originates from the selection rule derived for ideal suspended graphene [53]. The ultimate answer to this question will likely depend on specific experimental conditions, e.g. sample size, substrate coupling strength or strain. However, most of the studies agreed on a possibility of efficient control of the thermal conductivity in graphene nanoribbons via their geometrical shape, size and edge functionalization. The thermal conductivity can be made large for electronic device applications or small for the thermoelectric devices.

**Tuning Phonons via Interlayer Coupling**

In conventional semiconductor thin films the in-plane thermal conductivity decreases with decreasing thickness because the thermal transport in such structures is mostly limited by the phonon scattering from the film boundaries [84]. An opposite dependence can be observed in few-layer graphene (FLG) where the transport is limited mostly by the lattice anharmonicity [58]. Fig 4 shows that the thermal conductivity of the suspended uncapped FLG decreases with increasing number of the atomic layers *n* and approaches the bulk graphite limit. This trend was explained by considering the intrinsic quasi 2-D crystal properties described by the phonon Umklapp scattering [58]. As *n* in FLG increases – the additional phonon branches for the heat transfer appear, while, at the same time, more phase-space states become available for the phonon scattering. As a result the thermal conductivity decreases (see Fig. 4). The experimentally observed evolution of the thermal conductivity in FLG with *n* varying from 1 to *n*~4 [58] was also confirmed by independent theoretical investigations, which utilized different approaches [59-60, 85-86]. The thermal conductivity dependence on the FLG thickness can be entirely different for encased FLG, where thermal transport is limited by the acoustic phonon scattering from the top and bottom boundaries and disorder. An experimental study [51] found $K \approx 160$ W/mK for encased graphene at T=310 K. It increases to ~1000 W/mK for graphite films with the thickness of 8 nm. Thermal conduction in encased FLG was limited by the rough boundary scattering and disorder penetration through graphene. These results indicate that for efficient phonon engineering of thermal transport in FLG one needs to identify what transport regime – *intrinsic* vs. *extrinsic* – dominates the phonon transport in a given system.

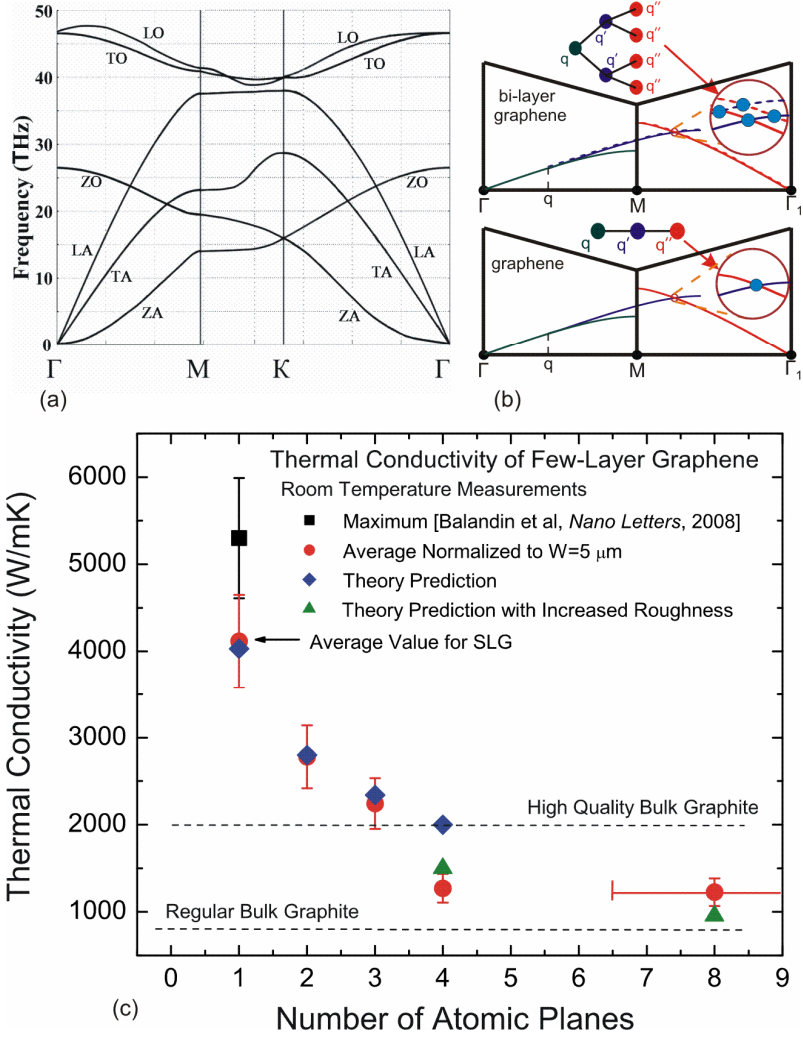

**Figure 4:** Phonon transport in graphene. (a) Phonon energy dispersion in a single layer graphene. In-plane longitudinal acoustic and optic (LA, LO), in-plane transverse acoustic and optic (TA, TO), and out-of-plane acoustic and optic (ZA, ZO) phonon branches are shown. (b) Diagram of three-phonon Umklapp scattering in graphene and bi-layer graphene (BLG), which shows that in BLG there are more states available for scattering owing to the increased number of phonon branches. (c) Measured thermal conductivity as a function of the number of atomic planes in few-layer graphene. The dashed straight lines indicate the range of bulk graphite thermal conductivities. The blue diamonds were obtained from the first-principles theory of thermal conduction in few-layer graphene based on the actual phonon dispersion and accounting for all allowed three-phonon Umklapp scattering channels. The green triangles are model calculations, which include extrinsic effects characteristic for thicker films. Figure is adopted from Ref. [58] with permission from the Nature Publishing Group.

**Controlling Phonons via Edge and Strain**

Since the average phonon MFP in graphene is about 800 nm near RT [39] the edge roughness scatterings affect the thermal conductivity even in the micrometer -wide graphene flakes. The latter substantially enhances the phonon engineering capabilities. It was theoretically found that the increase of the boundary specularity parameter from $p = 0.8$ to $p = 0.9$ changes the RT thermal conductivity of 5 μm – wide graphene flake by 30% from ~ 2200 W/mK to 3200 W/mK [40] (see Fig. 5). MD simulations suggested [63] that the thermal conductivity in rough-edged graphene ribbons is smaller than that in the smooth-edged ribbons by a factor of six. It was also shown that increasing of roughness from 0.1 nm to 1 nm leads to the drop in the thermal conductivity by an order of magnitude for the 10-nm wide graphene nanoribbon [64].

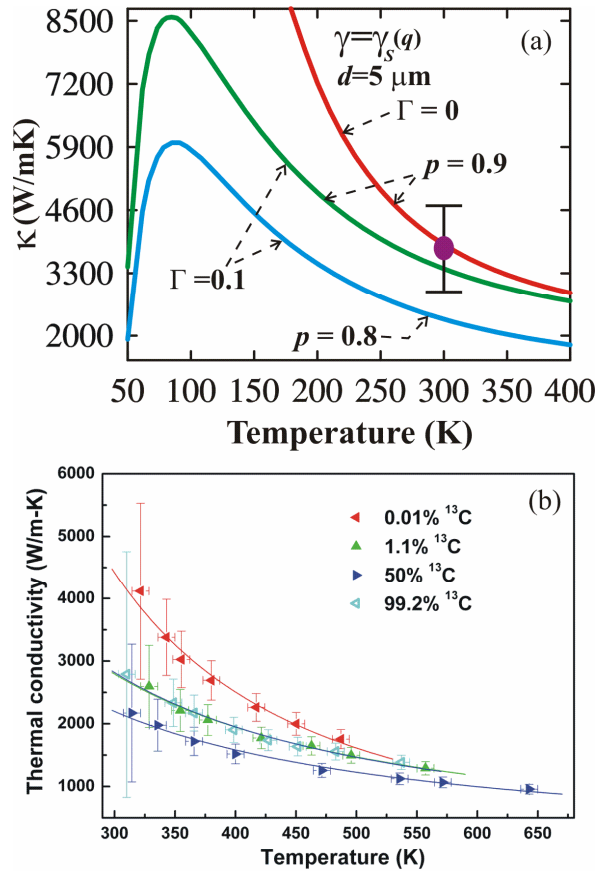

**Figure 5**: Isotope and defect engineering of phonon transport in graphene. (a) Thermal conductivity of graphene over a wide temperature range calculated for the graphene flake with the width of 5 μm and mode-dependent Gruneisen parameter. The results are obtained for two values of the specularity parameter $p=0.9$ and point-defect scattering strength $\Gamma$. An experimental data point after Refs. [38-39] is also shown for comparison. (b) Thermal conductivity of the suspended graphene film with $^{13}$C isotope concentrations of 0.01%, 1.1% (natural abundance), 50% and 99.2%, respectively, as a function of the temperature measured with the micro-Raman spectrometer. Note strong dependence of the thermal conductivity on the isotope concentration. Figure is reproduced from Refs. [40, 89] with permission from the American Physical Society and Nature Publishing Group.

The strain effects on the thermal conductivity of graphene nanoribbons were studied using MD simulations [87]. It was found that the thermal conductivity of graphene is very sensitive to tensile strain. Some of the strained graphene structures revealed higher thermal conductivity than that in graphene without strain [87]. The suggested strong edge and strain dependence of the thermal conductivity of graphene can explain some of discrepancies in the reported experimental values of the thermal conductivity. The suspended or supported graphene flakes and membranes are expected to have different strain field and different edge quality depending on the size and geometry of the suspected graphene sample as well as on the preparation technique. Changing strain field and edge roughness one can phonon engineer the thermal properties in graphene samples in a wide range of values at RT.

**Tuning Phonons in Graphene via Isotope and Defect Engineering**

Naturally occurring carbon materials are made up of two stable isotopes of $^{12}$C (~99%) and $^{13}$C (~1%). The relative concentrations of isotopes and crystal lattice defects can dramatic affect the thermal conductivity [40, 55, 88]. The BTE and MD simulation studies reported in Refs. [40, 55, 88] revealed the strong dependence of the thermal conductivity on the defect concentrations. It was shown that moderate increase in the defect concentration can decrease the thermal conductivity in graphene by a factor of five [40]. Other studies suggested that only 2 % of the vacancies can reduce the thermal conductivity of graphene by as much as a factor of five to ten [88]. An independent study found that the thermal conductivity can be reduced 1000-times at the vacancy concentration of ~9 % [55].

The first experimental study of the isotope effects on the phonon and thermal properties of graphene was reported just recently [89]. The isotopically modified graphene containing various percentage of $^{13}$C were synthesized by CVD technique [90-91]. The thermal conductivity of the isotopically pure $^{12}$C (0.01% $^{13}$C) graphene, measured by the optothermal Raman technique [37-39, 48, 58, 92], was higher than 4000 W/mK at temperature ~320 K, and more than a factor of two higher than the thermal conductivity of graphene sheets composed of a 50%-50% mixture of $^{12}$C and $^{13}$C (see Fig. 5). The evolution of thermal conductivity with the isotope content was attributed to the changes in the phonon – point defect scattering rate via the mass-difference term [89].

**Optical Phonons in Graphene and Few-Layer Graphene**

The fast progress in graphene field [93] was facilitated by the use of Raman spectroscopy as metrology tool [94]. Distinctive features in graphene and graphite Raman spectra – the zone center *G* peak and resonant *2D* band – allows one to count the number of atomic planes *n* in FLG as *n* changes from 1 to ~7 (see Fig. 6). For *n*>7 the spectrum becomes graphite like. The information about the number of atomic planes can be derived from the intensity ratio of *G* and *2D* peaks and deconvolution of the *2D* band on elemental peaks [94]. It was shown that the Raman metrology can be used for graphene on various substrates and different temperatures [95-96]. The ultraviolet Raman can provide additional metric for determining the number of atomic planes owing to different ratios of the *G* and *2D* peaks [97]. The disorder-induced *D* peak in Raman spectrum of graphene was instrumental in assessing concentration of defects and quality control of graphene synthesized by various techniques [98]. The knowledge of optical phonons and electron – phonon processes in graphene was important for development of graphene metrology techniques. From the other side, Raman spectroscopy was utilized to study the changes in graphene crystal lattice induced by the irradiation damage [99]. The sensitivity to temperature of the Raman peaks was instrumental for measuring the thermal conductivity of graphene [37-39, 48, 58, 91].

**Fine-Tuning Phonons in Rotationally Stacked Few-Layer Graphene**

There are two stable crystallographic stacking orders in bulk graphite: ABA Bernal stacking and ABC rhombohedral stacking. The staking orders of FLG strongly influence both electron [100-102] and phonon properties [101, 103-105]. It was demonstrated that ABA and ABC tri- and tetralayer graphene have different line shape and width of the Raman 2-D band [103]. Intriguing electron and phonon properties were found in the rotationally stacked or twisted FLG. When two graphene layers are placed on top of each other they can form the Moire pattern [101,105]. In this case one layer is rotated relative to another layer by an arbitrary angle. It was established that in the rotationally stacked bilayer a new peak *R'* appears in Raman spectra [105]. Depending on the rotational angle, the phonons with different wave vectors can manifest themselves in this peak. Investigation of Raman spectra of graphene, and misoriented bilayer graphenes led to the conclusion that in the misoriented bilayer graphene, the weak interaction between monolayers modifies the phonon dispersion curve in comparison with that in graphene while leaving the electronic band structure typically unaffected [101]. These conclusions are in agreement with more recent results pointing that in twisted graphene with rotational angle > $20^0$ the electronic properties are indistinguishable from those in graphene [102]. If the phonon properties of twisted FLG are also indistinguishable from those in graphene then it is possible that FLG atomic planes conduct heat as single-layer graphene. The latter may lead to extraordinary opportunities for thermal management with twisted FLG.

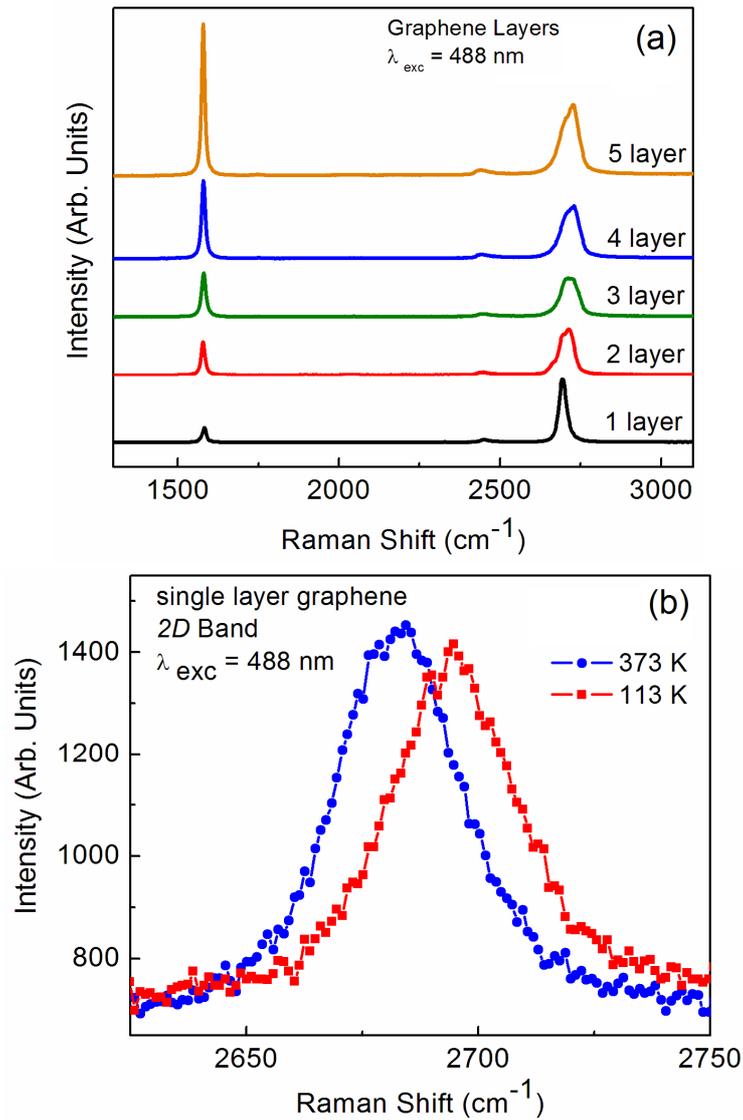

**Figure 6:** Optical phonons in graphene. (a) Evolution of Raman spectrum of graphene under the visible laser excitation at λ=488 nm as the number of atomic planes increases from *n*=1 to *n*=5. Raman spectra of graphene excited by 488 or 514 nm lasers are conventionally used for graphene identification and counting the number of graphene layers. (b) Raman spectrum showing 2D peak frequency at 113 and 373 K for a single layer graphene. Figure is reproduced from Refs. [96-97] with permission from the American Institute of Physics.

**PHONON THERMAL RECTIFICATION**

Phonon thermal rectification constitutes another interesting example of phonon engineering. Thermal rectification refers to a phenomenon predicted for materials systems where thermal conduction in one direction is better than in the opposite direction [106-109]. Controlling phonon transport by the structure asymmetry or lattice nonlinearity opens new possibilities, such as practical realization of the *thermal* diode. The proposed applications of the thermal rectifier vary from thermal management [107] to information processing [110]. One can distinguish several physical mechanisms leading to thermal rectifications and the corresponding material systems where they can be realized [56, 107, 111-114]. The first type is related to non-linear lattices with strong anharmonicity [109] or asymmetric mass or defect distributions [106,113]. There have been a number of theoretical predictions of thermal rectifications in such systems [107-108] but the experimental data is limited to experiments with the inhomogeneously mass-loaded CNTs [106]. Moreover, there are still questions on the theoretical interpretations of the rectification effects, particularly on the role of solitons [108]. The theories that predicted information processing with phonon rectifiers utilized idealized materials described by Hamiltonians with strong non-linearity, which are hard to find in nature. The second type of thermal rectification is more straightforward and feasible for practical applications. It is pertinent to the asymmetric structures operating in the ballistic or nearly-ballistic phonon transport regime where the acoustic phonon scattering from the boundaries is significant [111]. By modifying the geometry of the structure, introducing asymmetric edges or defects, one can make the phonon – boundary scattering stronger in one direction than in the opposite [111-112, 114].

**Two-Dimensional Graphene Phonon Rectifiers**

Graphene offers important advantages for investigation and possible practical applications of the thermal rectification effects owing to the long phonon MFP and stronger defect and grain boundary scattering than in the corresponding bulk. Considering that the average $\Lambda \sim 800$ nm at RT [39], the acoustic phonon transport in graphene is nearly ballistic up to micrometer length scale. Several groups have numerically studied the thermal rectification effects in GNR-based structures [56, 111-114]. The calculations were performed using either the non-equilibrium molecular dynamics (NEMD) or non-equilibrium Green functions (NEGF) methods. The considered graphene structures had nanoscale sizes due to the computational limitations. The nanometer size also ensures the ballistic phonon transport regime where the structure feature size $L<<\Lambda$.

The strength of thermal rectification is characterized by the degree of an inhomogeneity in the thermal flux via the coefficient $R \equiv (J_+ - J_-)/J_- \times 100\%$, where $J_+$ and $J_-$ are the heat fluxes along the opposite directions [107, 112-113]. Large $R$ values were predicted for triangle GNRs: $R \sim 120\%$ for $T = 150$ K and $R \sim 40\%$ at RT [56]. It was suggested that $R$ can reach 200% in the three-terminal graphene nanojunctions operating in the ballistic phonon transport regime [111]. Even larger $R$ of up to 400% was predicted for the few-nanometer-long trapezia-shaped GNRs [112]. Based on NEMD simulations it was suggested that the thermal rectification effect can be enhanced in bilayer graphene as compared to single layer graphene [114]. The increase in $R$ was attributed to the specifics of the interactions between the atomic planes. Thermal rectification effects were also predicted for the graphene-graphane interfaces owing to the asymmetry in Kapitza resistance [113]. While these computational studies still await their experimental confirmation, the results demonstrate potential of phonon engineering concept in 2D systems. Examples of the thermal applications of graphene and graphene ribbons can be found in the recent reviews [115-117].

**CONCLUSIONS**

We reviewed phonon properties in semiconductor nanostructures and two-dimensional graphene. Possibilities of controlled modification of the phonon dispersion and transport for achieving improved thermal and electronic properties – referred to as phonon engineering – were discussed in details. It was shown that quasi 2-D materials such as graphene and few-layer graphene offer many advantages for phonon engineering as compared to conventional nanostructures.

*Acknowledgements*

This work was supported, in part, by the National Science Foundation (NSF) projects US EECS-1128304, EECS-1124733 and EECS-1102074, by the US Office of Naval Research (ONR) through award N00014-10-1-0224, Semiconductor Research Corporation (SRC) and Defense Advanced Research Project Agency (DARPA) through FCRP Center on Functional Engineered Nano Architectonics (FENA), and DARPA-DMEA under agreement H94003-10-2-1003. DLN acknowledges the financial support through the Moldova State Project No. 11.817.05.10F. We acknowledge A. Askerov for his help in figure preparation.

**References**

[1] Stroscio, M.A., and Dutta, M., *Phonons in Nanostructures*, Cambridge University Press (2001), 1

[2] Balandin, A., and Wang, K. L. *J. Appl. Phys.* (1998) **84,** 6149

[3] Balandin, A., and Wang, K. L., *Phys. Rev*. B (1998) **58,** 1544

[4] Balandin, A. A., *J. Nanosci. Nanotech.* (2005) **5,** 1015

[5] Balandin, A. A. *et. al.*, *J. Nanoelect. Optoelect.* (2007) **2,** 140

[6] Klitsner, T., Pohl, R.O., *Phys. Rev. B* (1987) **36**, 6551

[7] Pokatilov, E.P., *et. al.*, *Appl. Phys. Lett.* (2004) **85**, 825

[8] Pokatilov, E.P., *et. al.*, *Phys. Rev. B* (2005) **72**, 113311

[9] Pokatilov, E.P., *et. al.*, *Superlatt. Microstruct.* (2005) **38,** 168

[10] Pokatilov, E.P., *et. al.*, *Superlatt. Microstruct.* (2003) **33,** 155

[11] Benisty, H., *et. al.*, *Phys. Rev. B* (1991) **44**, 10945

[12] Klimin, S.N., *et. al.*, *Phys. Stat. Sol. (b)* (1995) **190**, 441

[13] Pokatilov, E.P., *et. al.*, *Phys. Rev. B* (2008) **77**, 125328

[14] Rytov, S.M., *Sov. Phys. – Acoust.* (1956) **2**, 67

[15] Colvard, C., *et. al., Phys. Rev. B* (1985) **31**, 2080

[16] Bannov, N., *et. al., Phys. Stat. Sol. (b)* (1994) **183**, 131

[17] Nishiguchi, N., *et. al., J. Phys.: Condens. Matter.* (1997) **9**, 5751

[18] Svizhenko, A., *et. al*., *Phys. Rev. B* (1998) **57**, 4687

[19] Veliadis, J.V.D., *et. al., IEEE J. Quant. Elect.* (1996) **32**, 1155

[20] Zou, J., and Balandin, A., *J. Appl. Phys.* (2001) **89**, 2932

[21] Balandin, A.A., *Phys. Low-Dim. Structures* (2000) **5/6**, 73

[22] Ziman, J. M., *Electrons and Phonons: The Theory of Transport Phenomena in Solids,* Oxford University Press, New York (2001), 463

[23] Sakaki, H. *Jpn. J. Appl. Phys. (*1980) **19**, L735

[24] Fonoberov, V.A. and Balandin, A.A., *Nano Lett.* (2006) **6**, 2442

[25] Nika, D.L., *et. al.*, *Appl. Phys. Lett.* (2009) **93**, 173111

[26] Zincenco, N.D., *et. al.*, *J. Phys. Conf. Series* (2007) **92**, 012086

[27] Nika, D.L., *et. al.*, *J. Nanoelect. Optoelect.* (2009) **4**, 180

[28] Balandin, A.A. *et. al.*, in *Proceed. of XXII International Conference on Thermoelectrics ICT'22,* La Grand-Motte, France *(2003)*, IEEE 03TH8726, 399-402

[29] Lazarenkova, O.L., and Balandin, A.A., *Phys. Rev. B* (2002) **66**, 245319

[30] Balandin, A.A., and Lazarenkova, O.L., *Appl. Phys. Lett.* (2003) **82**, 415

[31] Nika, D.L., et. al., *Phys. Rev. B* **84**, 165415

[32] Pokatilov, E.P., et. al., *Appl. Phys. Lett.* (2006) **89**, 113508

[33] Pokatilov, E.P., et. al., *Appl. Phys. Lett.* (2006) **89**, 112110

[34] Pokatilov, E.P., et. al., *J. Appl. Phys.* (2007) **102**, 054304

[35] Hu, M., et. al., Nano Lett. (2011) 11, 618

[36] Wingert, M.C., *et. al., Nano Lett.* (2011) **11**, 5507

[37] Balandin, A.A., *Nature Mater.* (2011) **10**, 569

[38] Balandin. A.A., *et. al., Nano Letters* (2008) **8,** 902

[39] Ghosh, S., *et. al.*, *Appl. Phys. Lett*. (2008) **92,** 151911

[40] Nika, D.L., *et. al.*, *Phys. Rev. B* (2009), **79** 155413

[41] Nika, D.L., *et. al.*, *Appl. Phys. Lett*. (2009) **94,** 203103

[42] Lindsay, L., *et. al.*, *Phys. Rev. B* (2010) **82,** 115427

[43] Klemens, P.G., *J. Wide Bandgap Mater*. (2000) **7,** 332

[44] Klemens, P.G., *International J. Thermophysics* (2001) **22**, 265

[45] Ghosh, S., *et. al., New Journal of Physics* (2009) **11**, 095012

[46] Balandin, A.A., *et. al., ECS Transactions* (2010) **28**, 63

[47] Balandin, A.A., *et. al., Fullerenes, Nanotubes, and Carbon Nanostruct.* (2010) **18**, 1

[48] Cai, W., *et. al., Nano Lett.* (2010) **10**, 1645

[49] Jauregui, L.A., *et. al., ECS Transactions* (2010) **28**, 73

[50] Seol, J.H., *et. al. Science* (2010) **328,** 213

[51] Jang, W., *et. al. Nano Lett.* (2010) **10**, 3909

[52] Faugeras, C., *et. al.*, *ACS Nano* (2010) **4**, 1889

[53] Lindsay, L., *et. al.*, *Phys. Rev. B* (2010) **82**, 161402

[54] Qiu, B., Ruan, X., arXiv:111.4613v1 (2011)

[55] Zhang, H., *et. al., Phys. Rev. B* (2011) **84,** 115460

[56] Hu, J., *et. al.*, *Nano Lett.* (2009) **9**, 2730

[57] Haskins, J., *et. al.*, *ACS Nano* (2011) **5**, 3779

[58] Ghosh, S., *et. al.*, *Nature Mater.* (2010) **9**, 555

[59] Lindsay, L., *et. al.*, *Phys. Rev. B* (2011) **83**, 235428

[60] Singh, D., *et. al.*, *J. Appl. Phys.* (2011) **110**, 044317

[61] Wei, Z., *et. al., Carbon* (2011) **49**, 2653

[62] Murali, R., *et. al., Appl. Phys. Lett.* (2009) **94**, 243114

[63] Evans, W.J., et. al., *Appl. Phys. Lett.* (2010) **96**, 203112

[64] Aksamija, Z., and Knezevic, I., *Appl. Phys. Lett.* (2011) **98**, 141919

[65] Mak, K.F., *et. al., Phys. Rev. Lett.* (2011) **106**, 046401

[66] Munoz, E., *et. al., Nano Lett.* (2010) **10**, 1652

[67] Savin, A.V., *et. al., Phys. Rev. B* (2010) **82**, 195422

[68] Jiang. J.-W., *et. al. Phys. Rev. B* (2009) **79**, 205418

[69] Guo, Z., et. al., *Appl. Phys. Lett.* (2009) **95**, 163103

[70] Huang, Z., *et. al., J. Appl. Phys.* (2010) **108**, 094319

[71] Zhai, X., Jin, G., *EPL* (2011) **96**, 16002

[72] Jinag, J.-W., *et. al., Appl. Phys. Lett.* (2011) **98** 113114

[73] Ouyang, T., *et. al., EPL* (2009) **88**, 28002

[74] Droth, M., Burkard, G., *Phys. Rev. B* (2011) **84**, 155404

[75] Qian, J., *et. al., Superlatt. Microstruct.* (2009) **46**, 881

[76] Mounet, N., and Marzari, N., *Phys. Rev. B* (2005) **71**, 205214

[77] Wirtz, L., and Rubio, A., *Solid State Communications* (2004) **131**, 141

[78] Yan, J.-A., *et.al.*, *Phys. Rev. B* (2008) **77**, 125401

[79] Perebeinos, V. and Tersoff, J., *Phys. Rev. B* (2009) **79**, 241409(R).

[80] Nika, D.L., *et. al., Phys. Stat. Sol. (b)* (2011) **248**, 2609

[81] Lindsay, L., Broido, D., *Phys. Rev. B* (2010) **81**, 205441

[82] Maultzsch, J., *et. al., Phys. Rev. Lett.* (2004) **92**, 075501

[83] Mohr, M., *et. al., Phys. Rev. B* (2007) **76**, 035439

[84] Liu, W., and Asheghi, M., *J. Heat Transfer* (2006) **128**, 75

[85] Zhong, W.R. et. al., *Appl. Phys. Lett.* (2011) **98**, 113107

[86] Saito, K., and Dhar, A., *Phys. Rev. Lett.* (2010) **104**, 040601

[87] Wei, N., *et. al., Nanotechnology* (2011) **22**, 105705

[88] Hao, F., *et. al., Appl. Phys. Lett.* (2011) **99**, 041901

[89] Chen, S., *et. al., Nature Matert.* (2012) doi:10.1038/nmat3207

[90] Li, X., *et. al., J. Am. Chem. Soc.* (2011) **133**, 2816

[91] Li, X. S., *et. al., Nano Lett.* (2009) **9**, 4268

[92] Chen, S., *et. al. ACS Nano* (2011) **5**, 321

[93] Meyer, J.C., *et. al., Nature* (2007) **446**, 60

[94] Ferrari, A.C., *et. al., Phys. Rev. Lett.* (2006) **97**, 187401

[95] Calizo, I., *et. al., Nano Lett.* (2007) **7**, 2645

[96] Calizo, I., *et. al., Appl. Phys. Lett.* (2007) **91**, 071913

[97] Calizo, I., *et. al., J. Appl. Phys.* (2009) **106**, 043509

[98] Ioffe, A.F., *Semiconductor Thermoelements and Thermal Cooling,* Nauka (1956)

[99] Teweldebrhan, D., and Balandin, A.A., *Appl. Phys. Lett.* (2009) **94**, 013101

[100] Mak, K. F., *et. al., Phys. Rev. Lett.* (2010) **104**, 176404

[101] Poncharal, P., *et. al., Phys. Rev. B* (2008) **78**, 113407

[102] Luican, A., *et. al., Phys. Rev. Lett.* (2011) **106**, 126802

[103] Lui, C. H., *et. al., Nano Lett.* (2011) **11**, 164

[104] Cong, C., *et. al., Nano Lett.* (2011) **5**, 8760

[105] Carozo, V., *et. al., Nano Lett.* (2011) **11**, 4527

[106] Chang, C.W., *et. al., Science* (2006) **314**, 1121

[107] Casati, G., *Nature Nanotech.* (2007) **2**, 23

[108] Chen, J., *et. al., J. Stat. Mech.* (2011) P10031

[109] Terraneo, M., *et. al., Phys. Rev. Lett.* (2002) **88**, 094302

[110] Lei, W., and Li, B., *Phys. Rev. Lett.* (2007) **99**, 177208

[111] Ouyang, T., *et. al., Phys. Rev. B* (2010) **82**, 245403

[112] Yang, N., *et. al., Appl. Phys. Lett.* (2009) **95**, 033107

[113] Rajabpour, A., *el. al., Appl. Phys. Lett.* (2011) **99**, 051917

[114] Zhang, G., and Zhang, H., *Nanoscale* (2011) **3**, 4604

[115] Nika, D.L., and Balandin, A.A., J. Phys.: Condens. Matter (2012) **24**, 233203

[116] Shahil, K.M.F., and Balandin, A.A., Solid State Communications (2012); http://dx.doi.org/10.1016/j.ssc.2012.04.034

[117] Yan, Z., Liu, G., Khan, J.M., Balandin, A.A., Nature Communications (2012) 3, 827; DOI: 10.1038/ncomms1828